\documentclass[aps,prd,nofootinbib,
nobibnotes,twocolumn,floatfix,superscriptaddress,
letterpaper,showpacs,preprintnumbers]{revtex4-1}
\pdfoutput=1

\usepackage{amssymb,bbold}
\usepackage{graphicx, rotating}
\usepackage{epstopdf}
\usepackage{epsfig}
\usepackage{latexsym}
\usepackage{multirow}
\usepackage{color}
\usepackage{slashed}
\usepackage{natbib}
\usepackage{amsmath}
\usepackage{amsfonts}
\usepackage{verbatim}

\newcommand{\ba}{\begin{eqnarray}}
\newcommand{\ea}{\end{eqnarray}}
\newcommand{\no}{\nonumber}

\begin{document}

\title[Light sterile neutrinos from a late phase transition]
{Light sterile neutrinos from a late phase transition}
\author{Luca Vecchi}
\email{vecchi@pd.infn.it}
\affiliation{SISSA, via Bonomea 265, 34136, Trieste, Italy}
\affiliation{Dipartimento di Fisica e Astronomia, Universit\`a di Padova, via Marzolo 8, I-35131 Padova, Italy}
\affiliation{INFN, Sezione di Padova, via Marzolo 8, I-35131 Padova, Italy}

\begin{abstract}

Light sterile neutrinos represent a well-motivated extension of the 3-neutrino paradigm. However, the impressive agreement between standard cosmology and data casts doubts on their existence. Here we present a class of scenarios that robustly avoids this tension. In these models the sterile neutrinos are light, chiral states of a new sector interacting with the Standard Model via the right-handed neutrino portal and, crucially, active-sterile neutrino oscillations require a phase transition in the hidden sector. We explore the hidden-couplings/critical-temperature plane and identify regions where several sterile neutrinos can be accommodated. A late phase transition is usually preferred and may also ward off a potential threat posed by the formation of topologically stable defects.

\end{abstract}

\maketitle

\section{Motivation}

Light sterile neutrinos with sizable mixing with ordinary neutrinos are one of the primary targets of experimental efforts looking for new phenomena in neutrino physics. The reason is both theoretical and phenomenological. Firstly, sterile neutrinos are arguably the most plausible extension of the standard 3-neutrino paradigm that can significantly affect oscillations and yet simultaneously hide in all other channels. Secondly, sterile neutrinos of masses $m_s\sim1$ eV and mixing angles with active neutrinos of order $\theta\sim0.1$ are motivated by ``anomalies" in short baseline experiments, notably LSND~\cite{Aguilar:2001ty} (see also MiniBooNE~\cite{Aguilar-Arevalo:2013pmq}), and provide a simple interpretation of the so-called gallium~\cite{Acero:2007su} and reactor~\cite{Mention:2011rk} anomalies. (We suggest~\cite{Giunti:2015wnd} for a recent review and more references.)

Unfortunately, a sizable active-sterile mixing is not always a viable option within minimal extensions of the Standard Model. Indeed, scenarios with {\emph{truly-sterile}} light neutrinos coupled to the gauge-singlet combination of the lepton and Higgs doublets tend to predict a too large density of radiation when the Universe was at temperatures of order MeV and below~\cite{Dolgov}; this modifies the abundance of primordial elements during Big Bang Nucleosynthesis (BBN) and impacts the Cosmic Microwave Background (CMB) as well as large scale structure formation in a way that appears to be in conflict with data~\cite{Ade:2015xua}. One can alleviate this tension assuming huge primordial neutrino asymmetries~\cite{1} or diluting the sterile abundance via late entropy production~\cite{2}.

A more promising alternative is perhaps to abandon the minimal framework and consider scenarios where the exotic neutrinos are {\emph{non-sterile}}, i.e. states of a more involved hidden sector (see for example~\cite{Abazajian:2012ys,Adhikari:2016bei} for earlier references). In such a framework the non-sterile neutrinos may enjoy hidden interactions that can potentially impact their production in the early Universe and evade the cosmological constraints. For example, a sufficiently large exotic coupling may suppress active-sterile oscillations via a mechanism analogous to the MSW effect, as recently discussed in~\cite{Hannestad:2013ana,Dasgupta:2013zpn,Archidiacono:2014nda, Mirizzi:2014ama,Cherry:2014xra,Forastieri:2015paa,Chu:2015ipa,Cherry:2016jol}.

In this paper we propose a qualitatively different way to suppress thermal production of the non-sterile population that is realized in scenarios with {\emph{chiral}} sterile neutrinos. The basic observation is that in such scenarios active-sterile oscillations could not have started until after the hidden sector underwent a phase transition. As a consequence, cosmology can naturally be standard provided the critical temperature of the hidden dynamics is sufficiently low. 

Our mechanism differs from that of~\cite{Hannestad:2013ana,Dasgupta:2013zpn,Archidiacono:2014nda, Mirizzi:2014ama,Cherry:2014xra,Forastieri:2015paa,Chu:2015ipa,Cherry:2016jol} in many respects. First, it relies on a new symmetry, rather than a matter effect. Second, it works in an orthogonal region of the parameter space, where the couplings do not need to be sizable. Finally, it leads to a different phenomenology, e.g. it allows free-streaming neutrinos at CMB times.

In Section~\ref{sec:non} we introduce our framework and present models with Dirac neutrinos. (The Majorana neutrino portal is analyzed in Appendix~\ref{Majo}.) Cosmology is discussed in Section~\ref{cosmo} whereas Section~\ref{sec:astro} is devoted to the assessment of a few astrophysical constraints. We conclude in Section~\ref{sec:disc}, where the connection between our work and the recent literature is also elucidated.

\section{The Dirac neutrino portal}
\label{sec:non}

We add to the Standard Model (SM) the following Lagrangian (we use a Weyl fermion notation)
\ba\label{frame}
\delta{\cal L}&=&N^\dagger i\bar\sigma^\mu\partial_\mu N-(y_aNhL+{\rm h.c.})\\\no&-&(y_sN\phi\nu_s+{\rm h.c.}),
\ea
where the appropriate contractions of the Lorentz, flavor, and gauge indices are understood. Here $N$ is a familiar right-handed neutrino, $h$ the SM Higgs doublet, and $L=(\nu_a,\ell^-)^t$ the lepton doublet; $\nu_s$ are our (chiral) sterile neutrinos and $\phi$ is a scalar of the hidden sector. $y_a,y_s$ are Yukawa couplings and the subscripts $a,s$ stand for ``active" and ``sterile", respectively. Note that $N$ has a vanishing mass and acts as a mediator between the SM and the hidden sector. This is what we will call {\emph{the Dirac neutrino portal}}. (In our notation, the popular models of truly-sterile neutrinos mentioned in the Introduction have a small non-zero mass for $N$ and $y_s=0$.)

The couplings in (\ref{frame}), and in particular the absence of a mass for $\nu_s$, are enforced by a hidden gauge symmetry, of which $\phi\nu_s$ is a complete singlet (analogously to $hL$ in the SM). We assume the associated gauge coupling $g_s$ is weak for simplicity; a generalization to scenarios with larger $g_s$ can easily be carried out.

Scenarios of the type (\ref{frame}) have been considered previously in the presence of a mass for $N$. Mirror world models predict  $y_a=y_s$ (a relation we will not impose), with $N$ a singlet under the mirror parity. As far as we can tell, this was first noted in~\cite{Foot:1991py} for the case of Majorana neutrinos, though not so apparent from that reference. Moreover, in~\cite{Chun:1995js} a similar model --- again with $m_N\neq0$ --- was proposed as a solution of the solar neutrino problem. Analog constructions emerge in the context of the inverse seesaw mechanism~\cite{Mohapatra:1986bd}. More recently, ref.~\cite{Pospelov:2011ha} assumed a heavy $N$ and charged $\nu_s$ under a baryonic force stronger than weak. 
In the present paper we will instead assume neutrinos are Dirac ($m_N=0$) and neglect the hidden gauge force.

After $h,\phi$ have acquired a vacuum expectation value the neutral fermions $\nu_a,\nu_s$ mix. The parametric dependence of the $\nu_s-\nu_a$ mixing angles is given by
\ba\label{mix}
\tan\theta\sim{\rm min}\left({\frac{y_a\langle h\rangle}{y_s\langle\phi\rangle}},{\frac{y_s\langle\phi\rangle}{y_a\langle h\rangle}}\right).
\ea
We emphasize that $\nu_a-\nu_s$ mixing turns on after both the electroweak and the exotic dynamics have gone through a phase transition, and specifically $\langle\phi\rangle\neq0$ is a necessary condition. This will be important in Section~\ref{cosmo}.

There are ${\rm min}(n_a+n_s,n_N)$ Dirac neutrinos and $|n_a+n_s-n_N|$ unpaired chiral modes, where $n_{N,a,s}$ denote the numbers of Weyl fermions of the corresponding type ($n_a=3$). The chiral modes are in $N$ (or $\nu_s,\nu_a$) when $n_N>n_a+n_s$ (or $n_N<n_a+n_s$), and their masses are forbidden by a chiral symmetry. The mostly-active massive neutrinos have masses set by $y_a\langle h\rangle$, with atmospheric data suggesting
\ba\label{ya}
{\rm max}(y_a)\sim10^{-13}. 
\ea
On the other hand, the mostly-sterile massive combinations have $m_s\sim y_s\langle\phi\rangle$. Our benchmark point is $m_s\sim1$ eV and $\sin\theta=0.1$, as motivated by short baseline experiments. This requires $\langle\phi\rangle\gtrsim1$ eV.

 The assumptions $m_N\ll y_a\langle h\rangle$ and $y_a\ll1$ may be motivated by appropriate gauge symmetries.~\footnote{To forbid $m_N$ one can assume $B-L$. Analogously, to explain the smallness of $y_a$ one may invoke a second (local) symmetry under which $N$ is charged that forces $y_a$ to arise via the exchange of super heavy fields, effectively as a dimension-5 interaction. } However, in this paper we will view them as phenomenological inputs and refrain from speculating about their UV origin. In fact, we interpret (\ref{ya}) as indication that the particles $N,\nu_s,\phi$ are very weakly-coupled to our world. This will be a recurring theme in this work. 

We postulate our hidden sector is characterized by a mass scale $m_s$ that is much smaller than the weak scale. This appears to be compatible with the working hypothesis of negligible couplings between the hidden sector and the SM. For instance, a scalar mass $m_\phi\sim m_sy_s/4\pi$ and self-coupling $\lambda_\phi\sim y_s^4/16\pi^2$ are reasonable expectations, and automatically lead to $\langle\phi\rangle\sim m_s/y_s$.

\section{Cosmology of the Dirac portal}
\label{cosmo}

\subsection{Early decoupling}
\label{Dirac1}

At early times the exchange of super-heavy fields presumably kept $N,\phi,\nu_s$ in thermal equilibrium with the SM. As the temperature of the plasma dropped below the decoupling temperature $T_{\rm dec}$, the SM and the $N,\phi,\nu_s$ system started to evolve independently. When $T<T_{\rm dec}$, renormalizable interactions between $N,\phi,\nu_s$ and the SM, including $\nu_a-\nu_s$ oscillations, may have led to a late re-coupling.

However, all interactions of $\nu_s,\phi,etc.$ in (\ref{frame}) rely on $N$ exchange, and according to (\ref{ya}) $N$ is so weakly-coupled to the SM that it could have never been thermalized by $y_aNhL$~\cite{Shapiro:1980rr,Dolgov:1980cq}. As a result, the exotic particles $N,\nu_s,\phi$ were never brought to equilibrium by (\ref{frame}). Similarly, we postulate that any renormalizable coupling beyond those in (\ref{frame}) is suppressed. For example, to avoid a possible thermalization via the Higgs portal $\lambda_{\phi h}|\phi|^2|h|^2$ it is sufficient to take $\lambda_{\phi h}\ll10^{-8}$, whereas a kinetic mixing with hyper-charge $\epsilon B^{\mu\nu}F_{\mu\nu}'$ is irrelevant for $\epsilon\ll1$~\footnote{The models we consider here have a small hidden gauge coupling $g_s\ll1$ and can easily achieve this, since already a naive 1-loop estimate $\epsilon\sim g'g_s/16\pi^2$ may be enough. Scenarios with a sizable $g_s$ however rest on a few implicit assumptions, like the absence of fields charged under both symmetries and the presence of other small couplings in the theory.} or provided the hidden gauge symmetry is non-abelian, in which case a kinetic mixing would not arise in the first place.

Under these generic conditions, re-coupling at $T<T_{\rm dec}$ could have been triggered only by $\nu_a-\nu_s$ oscillations. This is the subject to which we now turn.

\subsection{Re-coupling via active-sterile mixing}
\label{plane}

We assume that the early Universe reached a high enough temperature so that the sterile chiral symmetry was restored, 
$$
\left.\langle\phi\rangle\right|_{T>T_\phi}=0. 
$$
No active-sterile mixing was allowed at those times, see (\ref{mix}). As the Universe cooled down a phase transition within the hidden sector took place and, at the critical plasma temperature $T_\phi$, the scalar $\phi$ acquired a vacuum expectation value.~\footnote{Because the hidden sector and the SM have comparable temperatures we make no distinction between the plasma temperature $T_\phi$ and the corresponding hidden-sector temperature.} Thus, active-sterile oscillations, and with it a possible re-coupling of $N,\nu_s,\phi,etc.$ and the visible sector, first became possible when 
\ba
T<T_\phi.
\ea
(We emphasize that in general there is no obvious relation between $\langle\phi\rangle$ and $T_\phi$. Furthermore, the phase transition may have been continuous as well as first order. However, in what follows we will have in mind scenarios with a second order phase transition at $T_\phi\sim\langle\phi\rangle$, for definiteness. More general frameworks are described in Appendix~\ref{Majo}.)

\begin{figure}[t]
\begin{center}
\includegraphics[width=7cm]{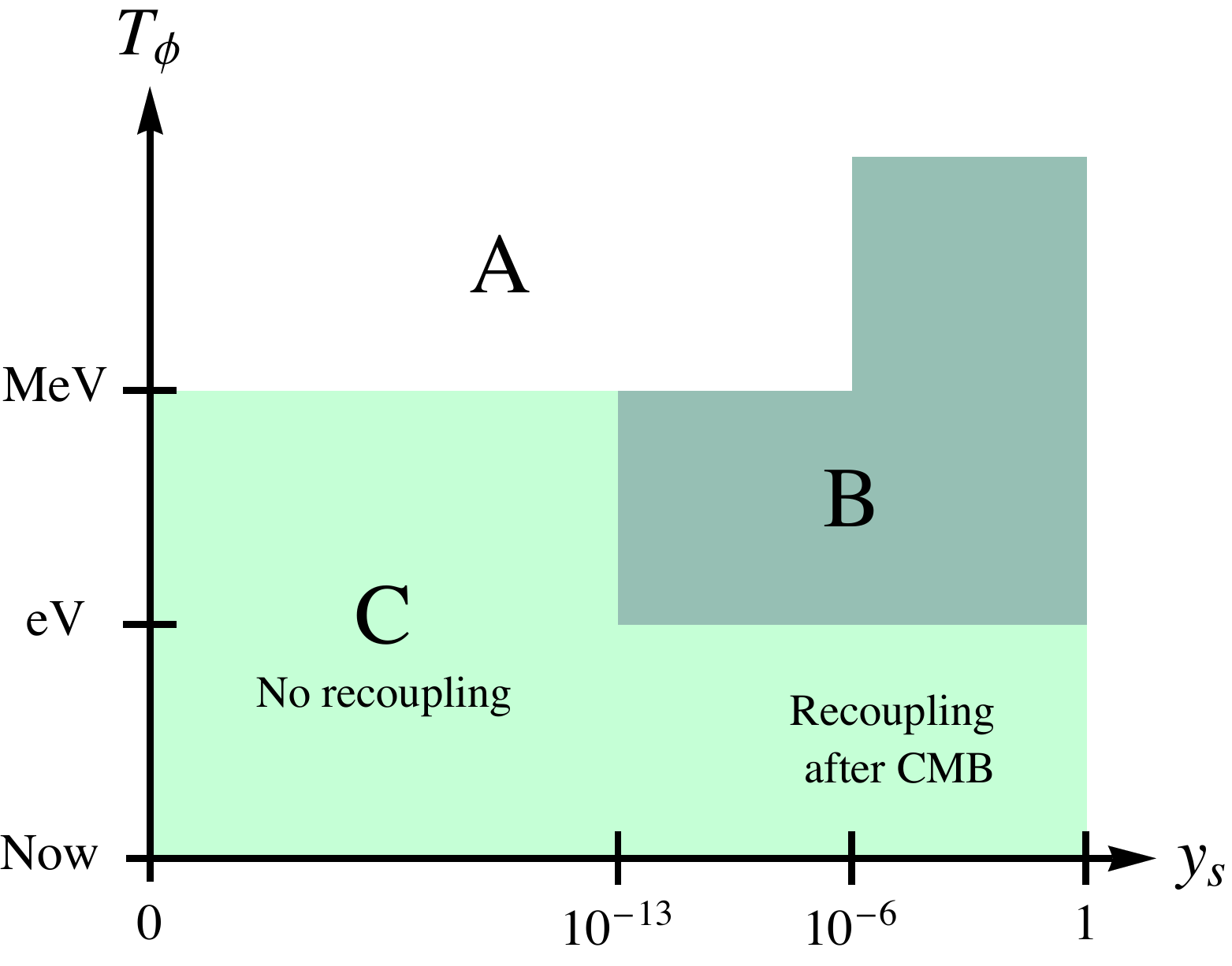}
\caption{Schematic plot of the 3 cosmologies: {\bf A} (truly-sterile), {\bf B} (collisional neutrinos during CMB), {\bf C} (no re-coupling until after CMB). 
}\label{Picture}
\end{center}
\end{figure}

Csmology at $T<T_{\rm dec}$ crucially depends on two unknown parameters: the critical temperature $T_\phi$ and the strength of the hidden interactions carried by $\nu_s$. In our models the most important hidden coupling turns out to be $y_s$ --- since $g_s$ is small by assumption. We distinguish 3 qualitatively different cosmologies, see regions {\bf A}, {\bf B}, {\bf C} in figure~\ref{Picture}.

\vspace{0.3cm}

{\bf{Region A: truly-sterile neutrinos.}} For $T_\phi>$ MeV sterile-active mixing is effective when the SM neutrinos are still in thermal equilibrium with the bath and one recovers a cosmology similar to that of the truly-sterile models. 

This however is true only at sufficiently small hidden coupling, see the white {\bf A} region in fig~\ref{Picture}. 

\vspace{0.3cm}

{\bf{Region {\bf B}: collisional neutrinos during CMB.}} As pointed out by refs.~\cite{Babu:1991at,Bento:2001xi} in the context of Majoron models (see also~\cite{Archidiacono:2014nda}), and more recently by~\cite{Hannestad:2013ana,Dasgupta:2013zpn,Mirizzi:2014ama,Cherry:2014xra,Forastieri:2015paa,Chu:2015ipa,Cherry:2016jol} for non-sterile neutrino models with gauge interactions, a qualitatively different picture is possible --- still when $T_\phi>$ MeV --- provided the hidden sector coupling is sufficiently large. The reason is that a large coupling generates a ``refractive index" for $\nu_s$ in the plasma that can suppress oscillations in a way similar to the MSW effect in matter. This mechanism can be efficient enough to forbid thermalization of the non-sterile neutrinos down to the BBN epoch and ensure $N_{\rm eff, BBN}\sim3$. To achieve this we need a thermal potential $V_s\sim y^2_sT$ larger than $\Delta m^2/T$ at $T\gtrsim1$ MeV. The resulting condition $y_s>m_s/$MeV, which reads $y_s>10^{-6}$ for $m_s=1$ eV, defines the upper part of the {\bf B} region. Here the deviation in the effective number of relativistic degrees of freedom compared to the SM (taken to be $N_{\rm eff, SM}=3$) at BBN is
\ba\label{Neff}
\Delta N_{\rm eff}=\left(n_{F}+\frac{4}{7}n_B\right)\left(\frac{g_{-}}{g_{+}}\right)^{4/3},
\ea
with $g_\pm$ the numbers of relativistic degrees of freedom at $T_{\rm dec}$ ($g_{+}\sim100$) and at neutrino decoupling ($g_{-}\sim10$), whereas $n_F=n_s+n_N$ and $n_B=n_\phi+n_{A'}$ (as long as $\phi,\nu_s$ and the hidden gauge field $A'$ are effectively massless). In deriving (\ref{Neff}) we assumed that the two sectors had the same temperature at $T>T_{\rm dec}$ and that the SM got hotter as a consequence of the QCD phase transition. In the absence of entropy production within the hidden sector, the ratio hidden/visible temperature -- $T_s/T_a=\left({g_{-}}/{g_{+}}\right)^{1/3}$ by conservation of entropy -- remained unchanged at later times.

There are other consequences of a large self-coupling, however. (In the present paragraph I will follow the arguments of~\cite{Mirizzi:2014ama,Cherry:2014xra,Forastieri:2015paa,Chu:2015ipa,Cherry:2016jol}.) As the temperature drops, the theories in the upper part of region {\bf B} of figure~\ref{Picture} entered a regime where active-sterile oscillations were very efficient and incoherent collisions of $\nu_s$ within the hidden sector --- parametrized by the rate $\Gamma_s$ --- so fast that 
\ba\label{coll}
\Gamma_s P(\nu_a\leftrightarrow\nu_s)> H 
\ea
well before the CMB epoch. The most relevant processes are $\nu\leftrightarrow N\phi$, and have rates of order $\Gamma_s\sim y_s^2 m^2/T$, where $m<T$ is the largest of the masses involved. Taking $m=1$ eV (and $\sin\theta=0.1$ as usual) we find that (\ref{coll}) holds whenever $y_s>10^{-13}$, interestingly close to (\ref{ya}). In a more optimistic scenario the heavy neutrinos have decayed and only the light ones participated to the 1 to 2 reactions, leading to a suppression of the collision rate. Nevertheless, we find that the condition (\ref{coll}) is met whenever the requirement $y_s>m_s/$MeV holds. We thus conclude that models in the upper part of region {\bf B} generally satisfy (\ref{coll}) at around $T\gtrsim1$ eV. As a result, the hidden and visible sectors equilibrated before CMB: the total neutrino number density got distributed among the active neutrinos and the exotic species while remaining the same. What this means is that active and sterile neutrinos formed a collisional plasma of relativistic states, rather than being free-streaming. CMB data may thus be used to constrain this region of parameter space, see~\cite{Mirizzi:2014ama,Cherry:2014xra,Forastieri:2015paa,Chu:2015ipa,Cherry:2016jol}.


In region {\bf B}, there exists another regime with suppressed thermalization of $\nu_s$ during and before the BBN epoch. This is realized when the phase transition in the hidden sector took place at eV $<T_\phi\lesssim1$ MeV. In this regime the mixing $\nu_a-\nu_s$ was {\emph{forbidden}} at BBN by a chiral symmetry of the hidden sector, irrespective of the size of the hidden sector couplings, and re-equilibration could not occur then. This mechanism to retain a standard BBN epoch is qualitatively different from the one considered in~\cite{Hannestad:2013ana,Dasgupta:2013zpn,Archidiacono:2014nda, Mirizzi:2014ama,Cherry:2014xra,Forastieri:2015paa,Chu:2015ipa,Cherry:2016jol}. In other words, in the lower part of region {\bf B} of Figure~\ref{Picture} the BBN epoch was automatically standard, but for a very different reason than the upper part. Nevertheless, the phenomenology at CMB would be quite similar if $y_s>10^{-13}$, because we would still end up with a non-free-streaming plasma of neutrinos in that case.

\vspace{0.3cm}

{\bf{Region {\bf C}: no re-coupling until after CMB.}} There is an interesting alternative, however. In scenarios with eV $<T_\phi\lesssim1$ MeV, active-sterile oscillations were precluded at and before BBN for any value of $y_s$. This allows us to consider the regime $y_s<10^{-13}$, as well. In these models, living in the upper part of region {\bf C} of fig~\ref{Picture}, BBN is standard and -- according to the discussion around eq.(\ref{coll}) -- neutrinos are free-streaming at CMB.

The absence of re-coupling until after CMB obviously extends to the lower part of the {\bf C} region, where oscillations were inhibited until $T_\phi<1$ eV. With such a low critical temperature any $y_s$ is automatically compatible with CMB data.

However, the impact of $y_s$ on large scale structures depends on the size of the coupling. With small $y_s$, neutrinos were free-streaming and structure formation proceeded as in the SM. In particular, the total neutrino mass $\sum_\nu m_\nu$ is automatically within the bounds of~\cite{Ade:2015xua}, since the energy density carried by the exotic neutrinos is small in our models. This corresponds to the part of {\bf C} region labeled ``No Re-coupling" in fig~\ref{Picture}. 

For larger $y_s$ there exists a temperature $T_*<$ eV above which neutrinos were collisional. This is the ``Re-coupling after CMB" region of fig~\ref{Picture}, which occurs in Region {\bf C} for $y_s$ larger than some model-dependent value. There is no obvious obstruction to this large $y_s$ limit~\cite{Mirizzi:2014ama}\cite{Chu:2015ipa}; however, such scenarios are harder to quantitatively assess because we currently lack an understanding of structure formation in the presence of a plasma of diffusing neutrinos.

In the entire {\bf C} region $N,\phi,\nu_s$ never re-coupled to the plasma between $1$ eV $ <T<T_{\rm dec}$, so the number of relativistic effective degrees of freedom is correctly approximated by~(\ref{Neff}), at BBN as well as CMB. For a representative set of $n_\phi, n_{A'}$ we find that $N_{\rm eff}$ is within the 1(2)-$\sigma$ bound of the CMB analysis of~\cite{Ade:2015xua} as long as $n_s+n_N$ is smaller than the numbers collected in table~\ref{tN}:

\begin{table}[h]
\begin{center}
\begin{tabular}{c||c|c|c} 
\rule[-.5em]{0pt}{1.2em} & $n_{A'}=2\times0$ & $n_{A'}=2\times1$ & $n_{A'}=2\times3$\\
\hline\hline
\rule[-.5em]{0pt}{1.6em}$n_\phi=1$ &${7(12)}$ & $6(10)$ & $3(8)$  \\
\hline
\rule[-.5em]{0pt}{1.6em}$n_\phi=2$ &$6(11)$ & ${5(10)}$ & $3(8)$  \\
\hline
\rule[-.5em]{0pt}{1.6em}$n_\phi=4$ &$5(10)$ & $4(9)$ & ${2(7)}$\\
\end{tabular}
\caption{\small 1(2)-$\sigma$ upper bound (from~\cite{Ade:2015xua}) on $n_s+n_N$. In all models we conservatively assumed that the exotic states are massless. The numbers in the diagonal refer to scenarios without or with a $U(1)$ or $SU(2)$ hidden gauge symmetry.
\label{tN}}
\end{center}
\end{table}

We conclude that scenarios in region {\bf C} of figure~\ref{Picture} can be consistent with Planck 2015 data~\cite{Ade:2015xua} (the constraint from the maximum value of the total neutrino mass quoted in~\cite{Ade:2015xua} is weaker than that from $N_{\rm eff, CMB}$). The hidden sector constitutes a negligible fraction of dark matter.

The mechanism that suppresses the sterile population in region {\bf C} is very robust -- because based on symmetry arguments -- and very different from that studied in~\cite{Hannestad:2013ana,Dasgupta:2013zpn,Archidiacono:2014nda, Mirizzi:2014ama,Cherry:2014xra,Forastieri:2015paa,Chu:2015ipa,Cherry:2016jol}: our mechanism forbids re-coupling rather than suppressing it, has different phenomenological consequences (neutrinos are not always collisional), and different parameters (e.g. no sizable couplings are needed).

\section{Astrophysics constraints}
\label{sec:astro}

In this section we discuss a number of constraints on non-sterile neutrinos. We rely only on generic features, like $\nu_a-\nu_s$ mixing and the existence of secret interactions for $\nu_s$, so our considerations should apply to a large class of scenarios.

\vspace{0.3cm}

{\bf{Neutrino propagation in the Cosmo.}} A flux of active neutrinos propagating with energy $E_\nu$ in a cosmic background of relic $\nu_a,\nu_s,N,\phi, A'$ can be depleted. The few anti-neutrinos from SN 1987a ($E_\nu\sim10$ MeV) --- and the $E_\nu\sim$ PeV neutrinos detected by IceCube --- can thus be used to set a limit on the hidden sterile interactions. 

There are many free parameters in our scenarios and we hence choose to be only qualitative here. We continue to assume the gauge interaction is weak and focus on the effect of $y_s$. For a more in depth discussion of the impact of secret interactions on the IceCube neutrinos we refer the reader to~\cite{Ioka:2014kca}, and~\cite{Cherry:2014xra} for a discussion of a model with vector interactions.

Neutrinos (and anti-neutrinos) are dominantly depleted via $\nu+{\rm bkd}\to\nu'+{\rm bkd}$, with $\nu$ a mostly-active neutrino (or anti-neutrino) state, bkd the cosmic background, and $\nu'$ a mostly $N,\nu_s$ state. We can set a conservative bound on $y_s$ assuming there exists a relic density $n_{\rm bkd}=100/$cm$^3$ of non-relativistic $\nu',\phi$ of mass $m_{\rm bkd}$ (we showed this is an overestimate in the previous section). In this case the scattering rate is proportional to $\sin^2\theta$, the smallest possible power of the mixing angle that allows depletion of the neutrino flux. The largest rates are found for $\nu\overline{\nu'}\to\phi\phi, \nu\phi\to\nu'\phi$, that have a log-enhanced cross section $\sigma\sim{\sin^2\theta y_s^4}\log(s/m_s^2)/{4\pi s}$ with $s\sim2 m_{\rm bkd}E_\nu$. Taking $\sin\theta=0.1$ and $m_{\rm bkd}=1$ eV, we find that the neutrino mean free path $\sim1/\sigma v_{\rm rel}n_{\rm bkd}$ is longer than the distance between the earth and SN 1987a ($\sim50$ Kpc) as soon as $y_s\lesssim10^{-2}$. Similar bounds were previously obtained in~\cite{Kolb:1987qy}. An independent limit on the high energy tail of $\sigma$ may be obtained if we knew the origin of the IceCube neutrinos. Yet, we should observe that as soon as $y_s\ll10^{-2}$ neutrinos can safely travel distances much longer than the current Universe.

\vspace{0.3cm}

{\bf{Supernovae cooling.}} Supernova (SN) cooling is affected by the exotic neutrino interactions, as well. Here we present a qualitative discussion of this effect.

We start observing that for $g_s\ll1$ the hidden gauge field $A'$ can freely exit the star. Because in this case $A'$ cannot acquire a thermal population, so does $\phi$. In such a regime, processes like $\nu_i+n\to \nu_j+n+A'/\phi$ can represent an important source of energy loss (here $\nu_i,\nu_j$ are mass eigenstates and $n$ is a neutron), in a way somewhat analogous to that induced by Majorons (see~\cite{Raffelt:1999tx} for earlier references and~\cite{Farzan:2002wx} for a recent discussion). $A'/\phi$ emission can only be efficient if the $\nu_i,\nu_j$ have a non-trivial overlap with the flavor $\nu_s$. However, in our scenario the active neutrinos (anti-neutrinos) that are produced in the inner regions have matter-suppressed couplings to $A'/\phi$.

Let us be a bit more quantitative. An order of magnitude estimate of the energy released per unit time and volume via $A'/\phi$ emission may be given by 
\ba\label{emiss}
\frac{d^2E}{dVdt}\sim\sin^2\theta_{\rm SN}\frac{g_*^2}{4\pi}\mu^4 V_{\rm eff},
\ea
where $\theta_{\rm SN}$ is the mixing angle at the $A'/\phi$ emission point and $g_*=g_s (y_s)$ for $A' (\phi)$, respectively. In the above expression we conservatively assumed that the relevant temperature and densities are both set by the chemical potential $\mu\sim100$ MeV, and neglected factors of order unity. $V_{\rm eff}\sim G_F\mu^3$ plays the role of an ``effective mass"  of the active neutrinos --- generated by interactions with the medium. The coherent decay $\nu_i\to\nu_j+A'/\phi$ dominates when $m_{A'},m_\phi\ll V_{\rm eff}$, whereas incoherent scattering typically leads to a much lower rate.  

Requiring that the energy loss within the neutrino-rich region (which we identify with the neutrino-sphere of radius $<100~{\rm km}$) be smaller than $\sim10^{53}$ erg/s we obtain $\sin\theta_{\rm SN}g_*<10^{-9}$. Now, taking as representative values $\Delta m^2\sim1$ eV and $E\sim10$ MeV, we find $\sin\theta_{\rm SN}\sim\sin\theta\Delta m^2/V_{\rm eff}E\sim\sin\theta\times10^{-8}$. Therefore, recalling that the emitted $A'/\phi$ is weakly coupled ($g_*\ll1$), we conclude that no significant constraint is anticipated.

In the strong $g_s$ limit the vector field can potentially be trapped inside the star and, for $m_\phi\lesssim T_{\rm SN}$, thermalize $\phi$ as well. This would result in an effective restoration of the exotic gauge symmetry in the core, with a consequent suppression of the mixing. The latter regime deserves a more dedicated study.

\vspace{0.3cm}

{\bf{Topological defects.}} Kibble pointed out that a phase transition characterized by the symmetry breaking patter ${\cal G}\to{\cal H}$ may result in the formation of stable topological field configurations in the early Universe if the manifold ${\cal M}={\cal G}/{\cal H}$ has non-trivial homotopy classes~\cite{Kibble:1976sj}. These defects are constrained by astrophysics and limit the parameter space of hidden sectors with phase transitions (including our models and those of~\cite{Hannestad:2013ana,Dasgupta:2013zpn,Archidiacono:2014nda, Mirizzi:2014ama,Cherry:2014xra,Forastieri:2015paa,Chu:2015ipa,Cherry:2016jol}), as we will now argue.

Following Zurek~\cite{Zurek}, one finds that in a second order transition at the temperature $T_\phi$, the typical correlation length $\xi$ in the early Universe is of order~\cite{Murayama:2009nj}
\ba\label{corre}
\xi=\frac{C}{T_\phi}\left(\frac{T_\phi}{H(T_{\phi})}\right)^{\gamma},
\ea
where $H(T)$ is the Hubble scale. In a weakly-coupled sigma model with quartic coupling $\lambda_\phi$ we have $\gamma=1/3$ and $C\sim10/\lambda_\phi^{1/3}$ (see e.g.~\cite{Chesler:2014gya}). We will take these as prototypical values and not attempt an estimate of $\xi$ for first order transitions (where $\xi$ would be set by the typical distance between bubbles of true vacuum).

$\xi$ determines the characteristic length scale of the defect; its energy density depends on the type of topological object and is of order $C_d(f\xi)^{d}f/\xi^3$, for $d=0,1,$ or $2$ depending on whether the defect is a monopole, string, or a domain wall. Ignoring a log dependence on the couplings: $C_0\sim1/g_s$ and $C_{1,2}\sim1$. $f$ sets the scale of the scalar vacuum expectation value and will be taken to be of order $f\sim T_\phi$. After being produced, the defects evolve in the expanding Universe, losing energy through scattering with the plasma and with other defects (emission of gravitational waves is negligible for $f\ll M_{\rm Planck}$). If we conservatively neglect the latter effects, their energy density will decrease with the scaling factor $R$ as $1/R^{3-d}$.

As a minimal constraint on these scenarios we demand that the energy density stored in the topological configuration be smaller than the vacuum energy $\rho_{\rm CC}\sim(2\times10^{-3}~{\rm eV})^4$ at temperatures $T\sim T_0$ of order the present epoch, $T_0=2.7$ K. The latter condition translates into an upper bound on the critical temperature:
\ba\label{bounds}
T_\phi\lesssim\sqrt{g_s/\lambda_\phi}~{\rm PeV}&~~~&{\rm monopole}~(d=0)\\\no
T_\phi\lesssim\lambda_\phi^{-1/4}~{\rm MeV}&~~~&{\rm string}~(d=1)\\\no
T_\phi\lesssim10\,\lambda_\phi^{-1/10}~{\rm eV}&~~~&{\rm domain~wall}~(d=2).
\ea
Because no accurate determination of the initial density and the subsequent evolution of the defects has been carried out, eq.(\ref{bounds}) should be taken as a crude, order of magnitude estimate of the actual bounds. Nevertheless, our result allows us to draw some qualitative conclusion. First, the formation of domain walls better be avoided in models with an early phase transition. Second, stable strings do not pose any serious threat as long as the phase transition occurred after the BBN epoch. Interestingly, this is precisely the new regime we motivated and discussed in Section~\ref{plane}.~\footnote{Anisotropies in the CMB due to cosmic strings are controlled by the ratio $f^2/M_{\rm Planck}^2$ and are undetectable for the range of parameters we are interested in.}

\section{Discussion}
\label{sec:disc}

On purely theoretical grounds, chiral sterile neutrinos may represent a natural byproduct of a hidden world with exotic gauge symmetries. The most economic way to couple such non-sterile neutrinos to the SM --- consistently with gauge invariance --- is via the right-handed neutrino portal~(\ref{frame}).

It is important to appreciate that the very existence of a gauge force for the non-sterile neutrinos has a number of non-trivial phenomenological implications, no matter how weak the hidden gauge coupling $g_s$ is. First, any realistic model with {\emph{chiral}} non-sterile neutrinos requires the presence of several light fermions to cancel the gauge anomalies in the hidden sector. For example, one can show that with a $U(1)$ hidden gauge symmetry, the minimal number of chiral fermions is $5$ --- not all of them could admit a coupling as in the second line of (\ref{frame}), however. For non-abelian groups that number further increases. Second, scenarios with Dirac and Majorana portals generically possess massive as well as light/massless states (see the discussion between (\ref{mix}) and (\ref{ya}) for the Dirac portal and Appendix~\ref{Majo} for Majorana portals). What this means in practice is that an explanation of the short baseline anomalies via non-sterile neutrinos would generically predict, on top of the familiar eV neutrinos, additional light fermions and the associated mass differences, some of which possibly comparable to those observed in solar and atmospheric data. This should be kept in mind when interpreting experimental results and performing global fits.

In this paper we argued that the cosmology of models of the type (\ref{frame}) can easily bypass the difficulties that characterize the more familiar scenarios of truly-sterile neutrinos (where $m_N\neq0, y_s=0$). In Section~\ref{cosmo} we showed that cosmology is a strong function of the hidden coupling $y_s$ and the critical temperature $T_\phi$ at which active-sterile mixing turned on --- i.e. at which $\phi$ acquired its vacuum expectation value. In this paper we identified the 3 qualitatively distinct regions in the $(y_s,T_\phi)$ plane, see figure~\ref{Picture}. 

In region {\bf A} the non-sterile neutrinos behave analogously to truly-sterile neutrinos and typically lead to $N_{\rm eff}>4$. Region {\bf B} has a standard BBN, but is constrained by our current understanding of structure formation and the CMB. On the other hand, in the domain {\bf C} the non-sterile neutrinos did not re-couple to the plasma until after CMB times, and our indirect probes of the BBN and CMB epochs can be easily consistent with the existence of several light exotic fermions. We focused on Dirac neutrino portals, but much of our analysis can be extended to Majorana portals, with the caveat emphasized in Appendix~\ref{Majo}.

Regions {\bf A}, {\bf B}, {\bf C} by no means exhaust all viable cosmologies of our scenarios (\ref{frame}), however. The basic working assumption of Section~\ref{cosmo} was that the hidden sector has been in thermal equilibrium with the SM at early times, when the temperature was above $T_{\rm dec}$. On the other hand, suppose the highest temperature in the early Universe has never reached $T_{\rm dec}$. In this case the hidden sector has not been in equilibrium with the SM until active-sterile oscillations turned on. Now, the temperature of the relic $\nu_s,\phi,N$ cannot be determined within our effective field theory formalism; however, their abundance should be irrelevant if after inflation the hidden sector temperature $T_s$ satisfied $T_\phi\ll T_s\ll T_{\rm SM}$. The resulting ``partially-reheated" scenario has an $N_{\rm eff}$ basically unaffected by the hidden world, provided the plasma temperature was below eV when $T_s<T_\phi$. Moreover, such scenarios can have a small collision rate $\Gamma_s\ll H$ irrespective of the actual strength of the hidden interactions, and may thus avoid equilibration of the neutrino species as well. This cosmology is quite plausible and modestly constrained by data.

To the best of our knowledge, the discussion of the different cosmologies in the $(y_s, T_\phi)$ plane of figure~\ref{Picture}, as well as the partially-reheated scenario meantioned in the previous paragraph, are novel results of this paper. Yet, the observation that a late phase transition can inhibit active-sterile oscillations was already made in~\cite{Chacko:2004cz}. However, the authors of~\cite{Chacko:2004cz} (see also~\cite{Mohapatra:2004uy} for a mirror-world scenario) discussed a fraction of the {\bf B} region with $10$ KeV $<T_\phi<1$ MeV, and analyzed a different class of models, proposed in~\cite{Chacko:2003dt}, where interactions analogous to those in (\ref{int}) re-coupled the hidden sector before CMB. On the other hand, our hidden world is completely decoupled, modulo $\nu_a-\nu_s$ mixing. The lower part of {\bf B} as well as region {\bf C} were not considered.

Our analysis also encompasses the framework studied in~\cite{Hannestad:2013ana,Dasgupta:2013zpn,Archidiacono:2014nda,Mirizzi:2014ama,Cherry:2014xra,Forastieri:2015paa,Chu:2015ipa,Cherry:2016jol}. These papers focused on the regime in which thermalization of the non-sterile neutrinos at BBN is suppressed by matter effects, but did not discuss the role of the phase transition, nor took into account the interactions (\ref{int}) (that are always present in those models). In our language, the work~\cite{Hannestad:2013ana,Dasgupta:2013zpn,Archidiacono:2014nda,Mirizzi:2014ama,Cherry:2014xra,Forastieri:2015paa,Chu:2015ipa,Cherry:2016jol} was confined to the upper part of {\bf B} with $T_\phi\gg$ MeV and implicitly assumed a heavy $\phi$. The rest of {\bf B} and region {\bf C} were not investigated. Yet, setting aside the fundamental origin behind the suppression of active-sterile oscillations at finite temperature, it should be stressed that at large exotic couplings the phenomenological signatures of our models is similar to those in~\cite{Hannestad:2013ana,Dasgupta:2013zpn,Archidiacono:2014nda,Mirizzi:2014ama,Cherry:2014xra,Forastieri:2015paa,Chu:2015ipa,Cherry:2016jol}: when $y_s$ is sufficiently large all models have collisional neutrinos before and/or after CMB. From a purely phenomenological perspective, therefore, it is the ``No Re-coupling" regime of region {\bf C} (i.e. the small $y_s$ limit) that uniquely characterizes the scenarios introduced here.

In Section~\ref{sec:astro} we presented a brief analysis of a number of astrophysics constraints on generic scenarios with non-sterile (Majorana or Dirac) neutrinos, including those of~\cite{Hannestad:2013ana,Dasgupta:2013zpn,Archidiacono:2014nda,Mirizzi:2014ama,Cherry:2014xra,Forastieri:2015paa,Chu:2015ipa,Cherry:2016jol}. The main original contribution there is a qualitative discussion of supernovae cooling and the impact of possible topological defects generated at the critical temperature $T_\phi$ via the Kibble-Zurek mechanism. While supernovae cooling does not seem to provide any relevant constraint, we found that topologically stable defects may be a concern for some of these scenarios. For instance, minimal non-sterile models with a $U(1)$ hidden gauge symmetry generically predict stable cosmic strings that may dominate the energy budged of our current Universe unless the exotic gauge symmetry was broken at a temperature $T_\phi\lesssim$ MeV. Interestingly, within this regime BBN is automatically standard irrespective of the strength of the exotic interactions (see Section~\ref{plane}). It should be stressed, however, that by promoting the exotic symmetry to, say, a $SU(n)$ fully broken by hidden scalars, all cosmological defects become unstable and the issue is completely obliterated. (As emphasized above, a non-abelian gauge symmetry also avoids a potential re-coupling induced by a kinetic mixing with hyper-charge.)

Our main conclusion is that realistic scenarios exist in which several light sterile neutrinos with sizable mixing to active neutrinos, such as those motivated by ``anomalies" in short baseline experiments, are consistent with cosmological data. For example, non-sterile neutrino models with Dirac portals in the {\bf C} region of figure~\ref{Picture} are plausible from a model-building point of view and, simultaneously, basically unconstrained by cosmology. Our work provides a first look at the phenomenology of these scenarios; a more systematic study is left for future work.


\begin{acknowledgments}

We thank I. Shoemaker for discussions on the bounds from SN 1987a and IceCube of Section~\ref{sec:astro}, and for sharing with us preliminary notes of~\cite{Cherry:2016jol} prior to publication. We are indebted to T. Battacharya and R. Gupta for their inquires on the nature of the phase transition at $T_\phi$, which eventually led to our discussion on topological defects. 
This work was supported by the ERC Advanced Grant no.267985 (DaMeSyFla) and the MIUR-FIRB grant RBFR12H1MW.

\end{acknowledgments}

\appendix

\section{The Majorana portal}
\label{Majo}

Suppose we add a mass term $-m_NNN/2+{\rm h.c.}$ to (\ref{frame}), and take $m_N\gg y_a\langle h\rangle$, $y_s\langle \phi\rangle$. In this framework, that we will refer to as the {\emph{Majorana neutrino portal}}, the low energy spectrum is composed of Majorana neutrinos.

Given generic mass matrices $y_a\langle h\rangle, y_s\langle \phi\rangle$ and $m_N\neq0$, the rank of the mass matrix in the $(\nu_a,\nu_s,N)$ basis is ${\cal R}=n_N+{\rm min}(n_a+n_s,n_N)$. $n_N$ of these neutrinos (mostly $N$) are heavy. The lighter ${\rm min}(n_a+n_s,n_N)$ states are dominantly $\nu_s,\nu_a$ with $\nu_s-\nu_a$ mixing angles of order (\ref{mix}). Some of them are massive, whereas $n_N+n_a+n_s-{\cal R}$ have vanishing tree-level masses. Masses for the latter arise at loop level because no symmetry prevents them.

Let us now turn to a brief discussion of the cosmology. We will follow Section~\ref{cosmo} and assume that the Higgs and gauge portals have no impact on our analysis. The main difference compared to the Dirac neutrino portal ($m_N=0$) is that additional constraints on the $\phi$ dynamics are required to maintain thermal decoupling below $T_{\rm dec}$.

As long as $y_{a,s}^2>m_N/M_*$ (with $M_*$ the reduced Planck mass) the interactions in (\ref{frame}) are large enough to thermalize $N$ and, with it, all the hidden sector at temperatures $T>m_N$. When the temperature of the plasma dropped below $T_{\rm dec}\sim m_N$ the two systems first decoupled. At $T$ much smaller than $m_N$ and the Higgs mass we can integrate $N$ out from (\ref{frame}) and, assuming $\phi$ is light, obtain (neglecting dimension-6 operators) 
\ba\label{int}
\delta{\cal L}_{\rm EFT}\supset \frac{m_a}{2}\nu_a\nu_a+ \frac{y_s^2}{2m_N}\nu_s\nu_s\phi\phi+y_*\phi\nu_a\nu_s+{\rm h.c.},~~~~~~~~
\ea
with $m_a=(y_a{\langle h\rangle})^2/{m_N}$ and $y_*=y_ay_s{\langle h\rangle}/{m_N}\sim \theta m_s/\langle\phi\rangle$. (All quantities are evaluated at zero temperature.) 

The last operator in Eq.~(\ref{int}) can re-couple the steriles to the plasma at $T\ll m_N$ via processes with physical or off-shell scalars. The important ones are $\phi\leftrightarrow\overline{\nu_a\nu_s}$ and $\nu_s\nu_a$ scattering, as well as their crossings. Consider first models with a light $\phi$. In this case the largest rates of the 1 to 2 processes scale as $y_*^2M^2/T$, where $M$ is the heaviest mass (either $m_\phi$ or $m_s$) and the lighter states are assumed to be massless with number density $\sim T^3$. These reactions turned on at $T_{\rm rc}\sim(y_*^2M^2M_*)^{1/3}$, usually well before CMB unless $\langle\phi\rangle$ is larger than a few TeV. (A much weaker constraint arises from the requirement that $\nu_s\nu_a$ scattering be out of equilibrium down to $T\sim1$ eV.) A smaller $\langle\phi\rangle$ is also possible, but at the expense of having a large $y_*$, i.e. collisional neutrinos during structure formation, as in~\cite{Chacko:2003dt,Chacko:2004cz}. In models with a heavy scalar, $\phi\leftrightarrow\overline{\nu_a\nu_s}$ ceased to be effective below temperatures of order $m_\phi$, whereas $\nu_s\nu_a$ scattering froze-out at $T_{\rm fo}\sim(m_\phi^4/y_*^4M_*)^{1/3}$. To decouple the hidden sector above the QCD phase transition as in Section~\ref{cosmo}, we impose $T_{\rm fo}>1$ GeV and obtain $\langle\phi\rangle>$ few KeV$\times({\rm GeV}/m_\phi)$.

From these considerations we conclude that Majorana portals satisfying the sigma-model relation $T_\phi\sim\langle\phi\rangle> m_\phi$ naturally live either in {\bf A} or the upper part of the  {\bf B} region. This is to be compared to the Dirac portals of Section~\ref{cosmo}, in which absence of re-coupling does not imply a lower bound on $\langle\phi\rangle$.

However, Majorana portals with $T_\phi<$ MeV may also be constructed if the symmetry-breaking sector is non-minimal, with $T_\phi,\langle\phi\rangle$ independent parameters. To appreciate how $T_\phi\ll\langle\phi\rangle$ or $T_\phi\gg\langle\phi\rangle$ may come about, consider a potential $V_\phi=m^2|\phi|^2+\lambda|\phi|^4$ and couple the scalar to $N_f$ fermions $\psi$ (and their conjugates $\overline{\psi}$) charged under an $SU(N_c)$ gauge theory via $y\overline{\psi}\psi\phi$. Classically, $\langle\phi\rangle=0$ is a stable solution. However, at a temperature $T_\phi$ the strong dynamics spontaneously breaks its chiral symmetry $\langle\overline{\psi}\psi\rangle=4\pi f^3$. (The phase transition can either be first order or continuous depending on $N_f,N_c$. For simplicity we may assume it is continuous and take $T_\phi\sim f$.) At the critical temperature the Yukawa coupling becomes a tadpole for $\phi$ and triggers $\langle \phi\rangle\neq0$. The expectation value of $\phi$ can be smaller or larger than $f\sim T_\phi$ depending on whether the ratio $m/f$ is large or small. More generally, the quantity $f$ may be communicated to the $\phi$ dynamics  via messenger fields that ``suppress" or ``amplify" the relevant mass scale, leading to either $\langle\phi\rangle\ll f$ or $\langle\phi\rangle\gg f$. The first option generically follows whenever the messengers are heavier than $f$. The hierarchy $\langle\phi\rangle\gg f$ can instead be obtained in models where the messenger is an approximate flat direction, in which case a small deformation of the potential (controlled by $f$) can result in a runaway behavior and a huge vacuum expectation value for the scalars.

Other scenarios with $T_\phi\ll\langle\phi\rangle$ are realized if the transition is first order. In that case $T_\phi$ should be identified with the temperature of bubble nucleation and the relation $T_\phi\ll\langle\phi\rangle$ is quite generic.~\footnote{The evolution of the $\nu_s$ density during a first order transition with $T_\phi\lesssim m_s$ is particularly amusing. At the nucleation temperature $T_\phi$ bubbles of the true vacuum would start to form inside the false vacuum. The non-sterile neutrinos are massive in the inside of the bubble, but massless outside. Hence, only an exponentially small fraction can penetrate. In practice, within this framework the population of exotic neutrinos gets expelled from our Universe.} However, a late first order transition may be problematic because of the necessity of avoiding a potential supercooling phase in which the free energy of the false vacuum dominates the expansion of the Universe. A second order transition might be preferable in this respect. Still, since the symmetry-breaking dynamics of the hidden sector is unknown --- only the portal to the visible world has been specified in~(\ref{frame}) --- no alternative should be discarded a priori.

In summary, Majorana portals with $T_\phi\ll1$ MeV (and larger $\langle\phi\rangle, m_\phi$) can be cooked up, but it is fair to say that region {\bf C} appears to be more generic for the Dirac portal.


\end{document}